\documentclass[aps,prd,floatfix,superscriptaddress,preprintnumbers,amsmath,amssymb]{revtex4}
\usepackage{graphicx,color,dcolumn,booktabs,bm}
\usepackage{longtable,lscape}
\usepackage{txfonts}
\usepackage{epsfig}
\usepackage{amssymb}
\usepackage{rotating}
\usepackage{cases}
\usepackage{color}
\usepackage{cases}
\usepackage{array}
\newcommand{\eq}{\begin{eqnarray}}
\newcommand{\en}{\end{eqnarray}}

\newcommand{\ra}{\rangle}

\begin{document}

\title{Structure and decays of hidden heavy pentaquarks}

\author{Thomas Gutsche}
\affiliation{Institut f\"ur Theoretische Physik,
Universit\"at T\"ubingen, \\
Kepler Center for Astro and Particle Physics,
Auf der Morgenstelle 14, D-72076 T\"ubingen, Germany}
\author{Valery E. Lyubovitskij}
\affiliation{Institut f\"ur Theoretische Physik,
Universit\"at T\"ubingen, \\
Kepler Center for Astro and Particle Physics,
Auf der Morgenstelle 14, D-72076 T\"ubingen, Germany}
\affiliation{Departamento de F\'\i sica y Centro Cient\'\i fico
Tecnol\'ogico de Valpara\'\i so-CCTVal, Universidad T\'ecnica
Federico Santa Mar\'\i a, Casilla 110-V, Valpara\'\i so, Chile}
\affiliation{Department of Physics, Tomsk State University,
634050 Tomsk, Russia}
\affiliation{Laboratory of Particle Physics, Tomsk Polytechnic University,
634050 Tomsk, Russia}

\date{\today}

\begin{abstract}

We study the hadronic molecular structure of hidden heavy pentaquarks --- 
new exotic states composed of charmed/bottom baryons and $D(D^*)/B(B^*)$ 
mesons. Based on the observation of three pentaquark candidates 
$P_c^+(4312)$, $P_c^+(4440)$, and $P_c^+(4457)$ by the LHCb Collaboration 
we consider the classification of possible flavor partners composed of 
charmed baryons and $D(D^*)$ mesons within the hadronic molecular approach. 
We extend this classification to the bottom sector. Using phenomenological 
Lagrangians we construct baryon-meson bound states governed by 
the Weinberg-Salam compositeness condition. As an application we consider 
strong two-body decays of the new exotic states into a light baryon and 
$V=J/\psi,\Upsilon$ or $P=\eta_c,\eta_b$ mesons. Results are presented 
in the heavy quark limit. 

\end{abstract}

\date{\today}
\maketitle
\section{Introduction}
\label{introduction} 

In 2015 the LHCb Collaboration reported on the observation of resonances in the 
$J/\psi p$ decay channel consistent with possible pentaquark states in the full 
reaction $\Lambda_b^0 \to J/\psi K^- p$ decay~\cite{Aaij:2015tga}. 
A model-dependent analysis of the invariant masses and angular distributions describing 
the $\Lambda_b^0$ decay lead to the claim of two pentaquark resonances, a broad state 
$P_c (4380)^+$ with mass $4380 \pm 8 \pm 29$ MeV and a width of $205\pm 18\pm 86$ MeV
and the narrower $P_c(4450)^+$ state with mass $4449.8 \pm 1.7 \pm 2.5$~MeV 
and width $39 \pm 5 \pm 19$ MeV. Soon after in Ref.~\cite{Aaij:2016phn} 
the LHCb Collaboration confirmed in a full amplitude analysis the consistency of data 
with the existence of the two exotic hadron structures $P_c (4380)^+$ and $P_c(4450)^+$. 
In a recent paper~\cite{Aaij:2019vzc} the LHCb Collaboration with the analysis of a much 
larger data sample confirmed the previously observed $P_c(4450)^+$ peak 
and resolved it into two narrow exotic baryon states $P_c(4440)^+$ and 
$P_c(4457)^+$. Furthermore, a narrow partner state $P_c(4312)^+$ 
has been claimed in~\cite{Aaij:2019vzc} while the existence of the $P_c (4380)^+$ can 
neither be confirmed nor excluded.
The conclusion drawn from this analysis was: 
(1) the minimal quark content of the $P_c$-states is $(duuc\bar c)$, 
(2) since the $P_c(4312)^+$, $P_c(4440)^+$, and $P_c(4457)^+$ 
are narrow and below the $\Sigma_c^+ \bar D^0$ and $\Sigma_c^+ \bar D^{* 0}$ 
thresholds, these states are strongly correlated with baryon-meson 
bound state structures.

Recently, the GlueX Collaboration at JLab~\cite{Ali:2019lzf} reported on the first measurement 
of the exclusive $J/\psi$ photoproduction cross section in the energy region from threshold 
up to $E_\gamma = 11.8$ GeV using a tagged photon beam. Such a measurement is extremely 
important since it provides a crucial check for theoretical approaches to the gluonic structure 
of the proton at high $x$, but also to possible structure interpretations of 
the LHCb pentaquarks.  
At this stage the GlueX Collaboration did not see any evidence for the LHCb pentaquarks and 
set model-dependent upper limits on their branching fractions ${\rm Br}(P_c^+ \to J/\psi p)$ 
with $4.6\%$ for $P_c(4312)^+$, $2.3\%$ for $P_c(4440)^+$, and $3.8\%$ for $P_c(4458)^+$ 
assuming spin-parity quantum numbers $J^P= \frac{3}{2}^-$ for each state. 

The observation of the LHCb Collaboration stimulated extensive theoretical 
studies of hidden pentaquark structures using different scenarios and 
frameworks (see, e.g., Ref.~\cite{Burns:2015dwa}-\cite{Peng:2019wys}  
and recent overviews in Refs.~\cite{Burns:2015dwa,Liu:2019zoy,Xiao:2019mvs}). 
In particular, the composite structure of the new exotic states has been 
studied using coupled-channel dynamics~\cite{Roca:2015tea,He:2015cea}.  
An application of QCD sum rules to hidden charm pentaquark states 
has been done in~\cite{Wang:2015epa} using diquark-diquark-antiquark type 
interpolating currents, in~\cite{Chen:2016otp,Azizi:2016dhy,Chen:2019bip} 
with meson-baryon molecular type currents, and the currents in form of product 
of two color-octet clusters of three light quarks and charm-anticharm 
pair~\cite{Pimikov:2019dyr}. Double polarization observables in pentaquark photoproduction 
have been studied by JPAC Collaboration using reaction model in Ref.~\cite{Winney:2019edt}. 
Different types of potential models to explain the spectrum of LHCb pentaquarks has 
been developed in Refs.~\cite{Zhu:2015bba}-\cite{He:2019ify}:    
a diquark-triquark potential model~\cite{Zhu:2015bba,Ali:2019clg}, 
quark delocalization color screening potential model~\cite{Huang:2015uda}, 
nonrelativistic potential model~\cite{Meng:2019fan},  chiral quark model~\cite{Yang:2015bmv}, 
color flux-tube model based on a five-body confinement potential~\cite{Deng:2016rus}, 
constituent quark model~\cite{Park:2017jbn}, 
diquark model derived using gauge/string duality~\cite{Giannuzzi:2019esi}, 
potential model based the Cornell-like potential~\cite{Xu:2019fjt}, 
and a quasipotential Bethe-Salpeter equation approach~\cite{He:2019ify}. 
Using a simple phenomenological model based on the G\"ursey-Radicati mass formula 
was used to predict the masses of hidden charm and bottom pentaquarks 
in Ref.~\cite{Holma:2019lxe}. 
In Refs.~\cite{Meng:2019ilv} the hidden charm and bottom pentaquark states 
have been studied using chiral perturbation theory. 
Implications of $SU(3)$ flavor symmetry for heavy pentaquarks have been 
considered in Refs.~\cite{Li:2015gta} and~\cite{Santopinto:2016pkp}. 
A hadronic molecular approach based on a 
charmonium-nucleon structure of the hidden charm 
pentaquarks has been proposed in Ref.~\cite{Eides:2015dtr}. 
An effective field theoretical approach incorporating heavy-quark spin symmetry 
has been suggested in Ref.~\cite{Liu:2019tjn}. 
Using effective effective Lagrangian approach 
the production of the pentaquark states  
$P_c(4312)$, $P_c(4440)$, and $P_c(4457)$ has been 
investigated in Refs.~\cite{Wang:2019krd}. 
A framework based on an effective-range expansion and resonance 
compositeness relations has been discussed in Ref.~\cite{Guo:2019kdc}.   
Field-theoretical hadronic molecular approaches for heavy pentaquarks 
have been developed in Refs.~\cite{Xiao:2019mvs,Guo:2019fdo}. 
In Refs.~\cite{Guo:2015umn} the new LHCb peaks have been related to 
kinematical effects in the rescattering from $\chi_{c1}$ to $J/\psi p$. 
Ref.~\cite{Eides:2019tgv} proposed a hadrocharmonium pentaquark scenario 
for the new states discovered by the LHCb Collaboration. 
Ideas of light- and heavy-flavor symmetries and their manifestation 
in properties of heavy hidden pentaquarks have been discussed 
in Refs.~\cite{Voloshin:2019aut}-\cite{Peng:2019wys}. 
One should also note that for the identification of the 
hidden charm pentaquark states it is important to perform a complete 
analysis of the cascade decay $\Lambda_b \to \Lambda^*(\frac{1}{2}^-, 
\frac{3}{2}^{\pm})[\to pK^-] + J/\psi$ done for example in 
Ref.~\cite{Gutsche:2017wag}. 

The main ideas in the application of quantum field theory to bound states using 
their compositeness have originally been laid out and formulated 
in Refs.~\cite{Salam:1962ap}-\cite{Efimov:1993ei}. 
Reference~\cite{Salam:1962ap} contains 
the original application to the composite system of the deuteron - the canonical
example of a hadronic molecule (HM). A extensive set of descriptions of hadronic 
molecules in the context of exotic heavy hadrons have been pursued by us for quite 
some time~\cite{Faessler:2007gv}-\cite{Dong:2017gaw}.

The main building blocks and related evaluation strategy of the quantum field approach 
to bound states~\cite{Salam:1962ap}-\cite{Efimov:1993ei} and specifically for the 
HM~\cite{Salam:1962ap},\cite{Faessler:2007gv}-\cite{Dong:2017gaw} are as follows: 
(1) First, a phenomenological, manifestly Lorentz covariant and gauge invariant Lagrangian 
has to be set up, which describes the interaction of the bound state with its constituents. 
The bound state and constituents are described by standard local quantum field operators. 
The field operators of the constituents form the interpolating current with the corresponding 
quantum numbers of the bound state; (2) The coupling strength of the hadronic molecule to 
its constituents is determined by the compositeness condition 
$Z_{\rm HM} = 1 - \Pi_{\rm HM}' = 0$~\cite{Salam:1962ap}-\cite{Dong:2017gaw}.
The wave function renormalization constant $Z_{\rm HM}$ of the hadronic molecule 
${\rm HM}$ defines the matrix element (or overlap) between the physical and bare states of 
the HM. $\Pi_{\rm HM}'$ is the derivative of the HM mass operator generated by the interaction 
Lagrangian of the HM with its constituents. The condition $Z_{\rm HM} = 0$ means that 
the probability to find the HM as a bare state is always equals zero or, in other words, 
it is always fully dressed by its constituents. 
The compositeness condition provides an effective and self-consistent way to describe 
the coupling of the HM to its constituents; (3) Then, using interaction Lagrangians between 
the HM and its constituents one can construct the $S$-matrix operator and consistently generate 
matrix elements for hadronic processes involving the HM (represented by corresponding Feynman 
diagrams). In the evaluation of the Feynman diagrams the compositeness condition enables to 
avoid the problem of double counting. 

The main objective of the present paper is a self-consistent study of the hidden charm 
pentaquarks composed of charmed baryons and $D$ mesons in hadronic molecular picture based 
on the formalism proposed and developed in Refs.~\cite{Faessler:2007gv}-\cite{Dong:2017gaw}.
We present a classification of these exotic states and calculate their strong two-body decays 
in the heavy quark limit.  
A first consideration of some of these states in a similar approach has been done 
recently~\cite{Xiao:2019mvs}. In our numerical analysis we proceed as follows: 
first we derive the results for the helicity amplitudes and decay rates 
in terms of two model parameters (size parameter $\Lambda$ and 
$D$-wave coupling of pseudoscalar heavy quarkonia with vector heavy-light meson and pair 
of light and heavy-light baryons). Next, we use recent results of 
the GlueX Collaboration~\cite{Ali:2019lzf} on upper limit of the branching fraction of 
the $P_c(4457)^+$ pentaquark to derive upper limit on its 
size parameter, which describes the distribution of the constituents in the pentaquark state. 

The paper is structured as follows.
In Sec.~II we give a classification of hidden charm pentaquarks -- 
exotic states composed of charmed baryons and $D$ mesons 
and present details of our formalism to set up these exotic states 
as hadronic molecules. We  also discuss the extension to hidden bottom pentaquarks. 
In Sec.~III we focus on the calculation of strong two-decays of hidden charm pentaquarks. 
We present a derivation of the corresponding matrix elements and the discuss numerical results. 
Finally, we summarize the results of the paper.

\section{Hadronic molecular structure of hidden charm pentaquarks}  
\label{framework} 

For the spin-parity quantum numbers of the hidden charm pentaquark states 
we use one of the possible scenarios, which follows from the conjecture of 
the LHCb Collaboration~\cite{Aaij:2019vzc} and the classification of some 
theoretical approaches: 
$J^{P} = \frac{1}{2}^{-}$ for the $P_c(4312)^+$ and $P_c(4440)^+$ states  
and $J^{P} = \frac{3}{2}^{-}$ for the $P_c(4457)^+$ state. We also base the following 
procedure on a base on a $SU_f(3)$ classification of hidden charm pentaquarks proposed
in Ref.~\cite{Santopinto:2016pkp} and consider all 
together 8 hidden pentaquarks states 
coinciding with hadronic molecular states composed of 
charmed baryons and $D(D^*)$ mesons. In Table~\ref{tab:hadrons} 
we present the classification of $24 = 3 \times 8$ hidden charm pentaquarks composed 
by a single charm baryon and $D(D^*)$ mesons. For each pentaquark 
discovered by the LHCb Collaboration we propose the existence of 7 partner states, 
which are composed of charmed baryons $(\Sigma_c, \Xi_c^\prime, \Omega_c)$ and $D(D^*)$ mesons. 
We specify isospin and spin-parity $I, J^P$, the interpolating currents 
in terms of the constituent fields, 
the constituent threshold (sum of the masses of the constituents), 
mass (if available from the LHCb Collaboration~\cite{Aaij:2019vzc}, otherwise our conjecture 
as explained further on). For the case when the pentaquarks are mixed states of two 
components we indicate the constituent thresholds for both cases (the value for the second 
component is given in brackets). In our conjecture for the masses of the pentaquarks 
we suppose that the mass difference of 
two pentaquark states is roughly equal to the difference of the corresponding 
charm baryon + $D(D^*)$ meson thresholds. By analogy, one can 
also derive hidden bottom pentaquarks replacing 
$D(D^*) \to B(B^*)$ and the charmed baryons by the bottom one. 
Recently a five-flavor classification of hidden charm and bottom 
pentaquarks has also been proposed in Ref.~\cite{Peng:2019wys}. Our scheme is 
differed since we only use nonstrange $D(D^*)$ and $B(B^*)$ heavy-light 
mesons in the construction of hidden heavy pentaquarks. We construct the 
pentaquark Fock states as eigenstates of the isospin operator 
$|I, I_3\ra$ and give an expansion in terms of the corresponding Fock states 
using standard $SU(2)$ couplings: 
\eq 
|1/2, \pm 1/2\ra &=& 
  \pm \sqrt{2/3} \ \biggl[ |1, \pm 1\ra \oplus |1/2, \mp 1/2\ra \biggr] 
\,\mp\, \sqrt{1/3} \ \biggl[ |1, 0\ra \oplus |1/2, \pm 1/2\ra \biggr] 
\,, \nonumber\\
|1/2, \pm 1/2\ra &=& |1/2, \pm 1/2\ra \oplus |0, 0\ra 
\,, \quad\quad\quad  
|1, \pm 1\ra \ = \ |1/2, \pm 1/2\ra \oplus |1/2, \pm 1/2\ra 
\,, \\
|1(0), 0\ra &=& 
     \sqrt{1/2} \ \biggl[ |1/2, 1/2\ra \oplus |1/2, -1/2\ra \biggr] 
\,\pm\,\sqrt{1/2} \ \biggl[ |1/2, -1/2\ra \oplus |1/2, 1/2\ra \biggr]
\,. \nonumber
\en 

\begin{table}[htb]
\begin{center}
\caption{Classification of hidden charm pentaquarks 
composed by a single charm baryon and $D(D^*)$ mesons} 

\vspace*{.1cm}

\def\arraystretch{1.2}
\begin{tabular}{c|c|c|c|c|c}
\hline
\ \ Pentaquark \ \
& \ \ $I$ \ \
& \ \ $J^P$ \ \
& \ \ Interpolating current $J_P$ \ \
& \ \ Threshold (MeV) \ \
& \ \ Mass (MeV) \ \ \\
\hline
 \multicolumn{6}{c}{First Family $P_{c1}$} \\
\cline{1-6}
\hline
$P_c(4312)^+$        & $\frac{1}{2}$   & $\frac{1}{2}^-$
&  $\sqrt{\frac{2}{3}} \Sigma_c^{++} D^-                                        
  - \sqrt{\frac{1}{3}} \Sigma_c^+ \bar D^0$
&  4323.62 (4317.17)
& $4311.9 \pm 0.7^{+6.8}_{-0.6}$~\cite{Aaij:2019vzc} \\
\hline
$P_c(4312)^0$        & $\frac{1}{2}$   & $\frac{1}{2}^-$
& -$\sqrt{\frac{2}{3}} \Sigma_c^{0} \bar D^0                                    
  + \sqrt{\frac{1}{3}} \Sigma_c^+ D^-$
&  4318.58 (4322.55)
& $4312$ \\
\hline
$P_c^{s}(4435)^+$    & $1$           & $\frac{1}{2}^-$
& $\Xi_c^{'+} \bar D^0$
&  4442.23
& $4435$ \\
\hline
$P_c^{s}(4435)^0$    & $1$           & $\frac{1}{2}^-$
& $\frac{1}{\sqrt{2}} \, \Big(\Xi_c^{'+} D^- + \Xi_c^{'0} \bar D^0\Big)$
&  4447.05 (4443.63)
& $4435$ \\
\hline
$P_c^{s}(4435)^-$    & $1$           & $\frac{1}{2}^-$
& $\Xi_c^{'0} D^-$
&  4448.45
& $4435$ \\
\hline
$\tilde P_c^{s}(4420)^0$    & $0$           & $\frac{1}{2}^-$
& $\frac{1}{\sqrt{2}}\,\Big(\Xi_c^{'+} D^- - \Xi_c^{'0} \bar D^0\Big)$
&  4447.05 (4443.63)
& $4420$ \\
\hline
$P_c^{ss}(4554)^0$   & $\frac{1}{2}$      & $\frac{1}{2}^-$
& $\Omega_c^{0} \bar D^0$
&  4560.03
& $45545$ \\
\hline
$P_c^{ss}(4554)^-$   & $\frac{1}{2}$      & $\frac{1}{2}^-$
& $\Omega_c^{0} D^-$
&  4564.85
& $4554$ \\
\hline
 \multicolumn{6}{c}{Second Family $P_{c2}$} \\
\cline{1-6}
\hline
$P_c(4440)^+$        & $\frac{1}{2}$   & $\frac{1}{2}^-$
& $\gamma^\mu \gamma^5 \,                                                       
\Big(\sqrt{\frac{2}{3}}                                                         
\Sigma_c^{++} D^{*-}_\mu                                                        
-  \sqrt{\frac{1}{3}} \Sigma_c^+ \bar D^{*0}_\mu\Big)$
& 4464.23 (4459.75)
& $4440.3 \pm 1.3^{+4.1}_{-4.7}$~\cite{Aaij:2019vzc} \\
\hline
$P_c(4440)^0$        & $\frac{1}{2}$   & $\frac{1}{2}^-$
& - $\gamma^\mu \gamma^5 \,                                                     
\Big(\sqrt{\frac{2}{3}} \Sigma_c^{0} \bar D^{*0}_\mu                            
-  \sqrt{\frac{1}{3}} \Sigma_c^+ D^{*-}\Big)$
& 4460.60 (4463.16)
& $4440$ \\
\hline
$P_c^{s}(4560)^+$    & $1$           & $\frac{1}{2}^-$
& $\gamma^\mu \gamma^5 \, \Xi_c^{'+} \bar D^{*0}_\mu$
& 4584.25
& $4560$ \\
\hline
$P_c^{s}(4560)^0$    & $1$           & $\frac{1}{2}^-$
& $\frac{1}{\sqrt{2}} \gamma^\mu \gamma^5 \,                                    
\Big(\Xi_c^{'+} D^{*-}_\mu + \Xi_c^{'0} \bar D^{*0}_\mu\Big)$
& 4587.66 (4585.65)
& $4560$ \\
\hline
$P_c^{s}(4560)^-$    & $1$           & $\frac{1}{2}^-$
& $\gamma^\mu \gamma^5 \, \Xi_c^{'0} D^{*-}_\mu$
& 4589.06
& $4560$ \\
\hline
$\tilde P_c^{s}(4545)^0$    & $0$           & $\frac{1}{2}^-$
& $\frac{1}{\sqrt{2}} \,                                                        
\gamma^\mu \gamma^5 \,                                                          
\Big(\Xi_c^{'+} D^{*-}_\mu - \Xi_c^{'0} \bar D^{*0}_\mu\Big)$
& 4587.66 (4585.65)
& $4545$ \\
\hline
$P_c^{ss}(4678)^0$   & $\frac{1}{2}$      & $\frac{1}{2}^-$
& $\gamma^\mu \gamma^5 \,\Omega_c^{0} \bar D^{*0}_\mu$
&  4702.05
& $4678$ \\
\hline
$P_c^{ss}(4678)^-$   & $\frac{1}{2}$      & $\frac{1}{2}^-$
& $\gamma^\mu \gamma^5 \,\Omega_c^{0} D^{*-}_\mu$
&  4705.46
& $4678$ \\
\hline
 \multicolumn{6}{c}{Third Family $P_{c3}$} \\
\cline{1-6}
\hline
$P_c(4457)^+$        & $\frac{1}{2}$   & $\frac{3}{2}^-$
& $\sqrt{\frac{2}{3}}                                                           
\Sigma_c^{++} D^{*-}_\mu                                                        
-  \sqrt{\frac{1}{3}} \Sigma_c^+ \bar D^{*0}_\mu$
&  4464.23 (4459.75)
& $4457.3 \pm 0.6^{+4.1}_{-1.7}$~\cite{Aaij:2019vzc} \\
\hline
$P_c(4457)^0$        & $\frac{1}{2}$   & $\frac{3}{2}^-$
& - $\sqrt{\frac{2}{3}} \Sigma_c^{0} \bar D^{*0}_\mu                            
+  \sqrt{\frac{1}{3}} \Sigma_c^+ D^{*-}$
&  4460.60 (4463.16)
& $4457$ \\
\hline
$P_c^{s}(4575)^+$    & $1$           & $\frac{3}{2}^-$
& $\Xi_c^{'0} \bar D^{*0}_\mu$
&  4584.25
& $4575$ \\
\hline
$P_c^{s}(4575)^0$    & $1$           & $\frac{3}{2}^-$
& $\frac{1}{\sqrt{2}} \,                                                        
\Big(\Xi_c^{'+} D^{*-}_\mu + \Xi_c^{'0} \bar D^{*0}_\mu\Big)$
&  4587.66 (4585.65)
& $4575$ \\
\hline
$P_c^{s}(4575)^-$    & $1$           & $\frac{3}{2}^-$
& $\Xi_c^{'0} D^{*-}_\mu$
&  4589.06
& $4575$ \\
\hline
$\tilde P_c^{s}(4545)^0$    & $0$           & $\frac{3}{2}^-$
& $\frac{1}{\sqrt{2}} \,                                                        
\Big(\Xi_c^{'+} D^{*-}_\mu - \Xi_c^{'0} \bar D^{*0}_\mu\Big)$
&  4587.66 (4585.65)
& $4545$ \\
\hline
$P_c^{ss}(4695)^0$   & $\frac{1}{2}$      & $\frac{3}{2}^-$
& $\Omega_c^{0} \bar D^{*0}_\mu$
& 4702.05
& $4695$ \\
\hline
$P_c^{ss}(4695)^-$   & $\frac{1}{2}$      & $\frac{3}{2}^-$
& $\Omega_c^{0} D^{*-}_\mu$
& 4705.46
& $4695$ \\
\hline
\end{tabular}
\label{tab:hadrons}
\end{center}
\end{table}

\begin{table}[htb]
\begin{center}
\caption{Classification of hidden bottom pentaquarks 
composed of vva single bottom baryon and $B(B^*)$ mesons} 

\vspace*{.1cm}

\def\arraystretch{1.2}
\begin{tabular}{c|c|c|c|c|c}
\hline
\ \ Pentaquark \ \
& \ \ $I$ \ \
& \ \ $J^P$ \ \
& \ \ Interpolating current $J_P$ \ \
& \ \ Threshold (MeV) \ \
& \ \ Mass (MeV) \ \ \\
\hline
 \multicolumn{6}{c}{First Family $P_{b1}$} \\
\cline{1-6}
\hline
$P_b(11080)^+$        & $\frac{1}{2}$   & $\frac{1}{2}^-$
& $\sqrt{\frac{2}{3}} \Sigma_b^{+} B^0                                          
-  \sqrt{\frac{1}{3}} \Sigma_b^0 B^+$
&  11090.20 (11092.33)
&  11080 \\
\hline
$P_b(11080)^0$        & $\frac{1}{2}$   & $\frac{1}{2}^-$
& - $\sqrt{\frac{2}{3}} \Sigma_b^{-} B^+                                        
+  \sqrt{\frac{1}{3}} \Sigma_b^0 B^0$
&  11094.97 (11092.64)
&  11080 \\
\hline
$P_b^{s}(11215)^+$    & $1$           & $\frac{1}{2}^-$
& $\Xi_b^{'0} B^+$
&  11214.35
&  11215 \\
\hline
$P_b^{s}(11215)^0$    & $1$           & $\frac{1}{2}^-$
& $\frac{1}{\sqrt{2}} \, \Big(\Xi_b^{'0} B^0 + \Xi_b^{'-} B^+\Big)$
& 11214.66 (11214.35)
& 11215 \\
\hline
$P_b^{s}(11215)^-$    & $1$           & $\frac{1}{2}^-$
& $\Xi_b^{'-} B^0$
& 11214.72
& 11215 \\
\hline
$\tilde P_b^{s}(11200)^0$    & $0$           & $\frac{1}{2}^-$
& $\frac{1}{\sqrt{2}}\,\Big(\Xi_b^{'0} B^0 - \Xi_b^{'-} B^+\Big)$
& 11214.66 (11214.35)
& 11200 \\
\hline
$P_b^{ss}(11315)^0$   & $\frac{1}{2}$      & $\frac{1}{2}^-$
& $\Omega_b^{-} B^+$
&  11325.43
&  11315 \\
\hline
$P_b^{ss}(11315)^-$   & $\frac{1}{2}$      & $\frac{1}{2}^-$
& $\Omega_b^{-} B^0$
&  11325.74
&  11315 \\
\hline
 \multicolumn{6}{c}{Second Family $P_{b2}$} \\
\cline{1-6}
\hline
$P_b(11125)^+$        & $\frac{1}{2}$   & $\frac{1}{2}^-$
& $\gamma^\mu \gamma^5 \,                                                       
\Big(\sqrt{\frac{2}{3}}                                                         
\Sigma_b^{+} B^{*0}_\mu                                                         
-  \sqrt{\frac{1}{3}} \Sigma_b^0 B^{*+}_\mu\Big)$
& 11135.26 (11137.70)
& 11125 \\
\hline
$P_b(11125)^0$        & $\frac{1}{2}$   & $\frac{1}{2}^-$
& - $\gamma^\mu \gamma^5 \,                                                     
\Big(\sqrt{\frac{2}{3}} \Sigma_b^{-} B^{*+}_\mu                                 
- \sqrt{\frac{1}{3}} \Sigma_b^{0} B^{*0}\Big)$
& 11140.34 (11137.70)
& 11125 \\
\hline
$P_b^{s}(11250)^+$    & $1$           & $\frac{1}{2}^-$
& $\gamma^\mu \gamma^5 \, \Xi_b^{'0} B^{*+}_\mu$
& 11259.72
& 11250 \\
\hline
$P_b^{s}(11250)^0$    & $1$           & $\frac{1}{2}^-$
& $\frac{1}{\sqrt{2}} \gamma^\mu \gamma^5 \,                                    
\Big(\Xi_b^{'0} B^{*0}_\mu + \Xi_b^{'-} B^{*+}_\mu\Big)$
& 11259.72 (11259.72)
& 11250 \\
\hline
$P_b^{s}(11250)^-$    & $1$           & $\frac{1}{2}^-$
& $\gamma^\mu \gamma^5 \, \Xi_b^{'-} B^{*0}_\mu$
& 11259.72
& 11250 \\
\hline
$\tilde P_b^{s}(11235)^0$    & $0$           & $\frac{1}{2}^-$
& $\frac{1}{\sqrt{2}} \,                                                        
\gamma^\mu \gamma^5 \,                                                          
\Big(\Xi_b^{'0} B^{*0}_\mu - \Xi_b^{'-} B^{*+}_\mu\Big)$
& 11259.72 (11255.72)
& 11235 \\
\hline
$P_b^{ss}(11360)^0$   & $\frac{1}{2}$      & $\frac{1}{2}^-$
& $\gamma^\mu \gamma^5 \,\Omega_b^{-} B^{*+}_\mu$
& 11370.80
& 11360 \\
\hline
$P_b^{ss}(11360)^-$   & $\frac{1}{2}$      & $\frac{1}{2}^-$
& $\gamma^\mu \gamma^5 \,\Omega_b^{-} B^{*0}_\mu$
& 11370.80
& 11360\\
\hline
 \multicolumn{6}{c}{Third Family $P_{b3}$} \\
\cline{1-6}
\hline
$P_b(11130)^+$        & $\frac{1}{2}$   & $\frac{3}{2}^-$
& 
$\sqrt{\frac{2}{3}}                                                             
\Sigma_b^{+} B^{*0}_\mu                                                         
-  \sqrt{\frac{1}{3}} \Sigma_b^0 B^{*+}_\mu$
& 11135.26 (11137.70)
& 11130 \\
\hline 
$P_b(11130)^0$        & $\frac{1}{2}$   & $\frac{3}{2}^-$
& - $\sqrt{\frac{2}{3}} \Sigma_b^{-} B^{*+}_\mu                                 
+   \sqrt{\frac{1}{3}} \Sigma_b^{0} B^{*0}$
& 11140.34 (11137.70)
& 11130 \\
\hline
$P_b^{s}(11255)^+$    & $1$           & $\frac{3}{2}^-$
& $\Xi_b^{'0} B^{*+}_\mu$
& 11259.72
& 11255 \\
\hline
$P_b^{s}(11255)^0$    & $1$           & $\frac{3}{2}^-$
& $\frac{1}{\sqrt{2}} \, 
\Big(\Xi_b^{'0} B^{*0}_\mu + \Xi_b^{'-} B^{*+}_\mu\Big)$
& 11259.72 (11259.72)
& 11255 \\
\hline
$P_b^{s}(11255)^-$    & $1$           & $\frac{3}{2}^-$
& $\gamma^\mu \gamma^5 \, \Xi_b^{'-} B^{*0}_\mu$
& 11259.72
& 11255 \\
\hline
$\tilde P_b^{s}(11240)^0$    & $0$           & $\frac{3}{2}^-$
& $\frac{1}{\sqrt{2}} \,                                                        
\Big(\Xi_b^{'0} B^{*0}_\mu - \Xi_b^{'-} B^{*+}_\mu\Big)$
& 11259.72 (11259.72)
& 11240 \\
\hline
$P_b^{ss}(11365)^0$   & $\frac{1}{2}$      & $\frac{3}{2}^-$
& $\Omega_b^{-} B^{*+}_\mu$
& 11370.80
& 11365 \\
\hline
$P_b^{ss}(11365)^-$   & $\frac{1}{2}$      & $\frac{3}{2}^-$
& $\Omega_b^{-} B^{*0}_\mu$
& 11370.80
& 11365\\
\hline
\end{tabular}
\label{tab:hadronsb}
\end{center}
\end{table}

We next describe the hadronic molecular structure of the pentaquarks using 
phenomenological Lagrangians which involve the interpolating currents 
presented in Table~\ref{tab:hadrons}. Note that the three states 
$P_c(4312)^+$, $P_c(4440)^+$, and $P_c(4457)^+$ have already been considered 
in Ref.~\cite{Xiao:2019mvs}. For all 24 pentaquark states 
the corresponding Lagrangians look as 
\eq\label{Lagr_Pc} 
{\cal L}_{P_c}(x) &=&  g_{P_{c1}} \, \bar P_{c1}(x) \, J_{P_{c1}}(x) 
                 \,+\, g_{P_{c2}} \, \bar P_{c2}(x) \, J_{P_{c2}}(x)
                 \,+\, g_{P_{c3}} \, \bar P_{c3,\mu}(x) \, J_{P_{c3}}^\mu(x) 
+ {\rm H.c.}\,, 
\en                                                             
where $P_{c1}$, $P_{c2}$, and $P_{c3,\mu}$ are the hidden 
charm pentaquark fields belonging to the first, second, and 
third family, respectively, 
$J_{P_{c1}}(x)$, $J_{P_{c2}}(x)$, and $J_{P_{c3}}^\mu(x)$ 
are the nonlocal extension of the 
currents from Table~\ref{tab:hadrons}, $g_{P_{ci}}$ $(i=1,2,3)$ 
is the coupling constant. 
The nonlocal pentaquark currents are written as (here flavor indices 
are suppressed)
\eq
J_{P_{c1}}(x) &=& \int d^4y \, \Phi_{P_{c1}}(y^2) \ 
H_c(x+\omega_Dy) \bar D(x-\omega_{H_c}y) \,, \label{Curr_Pc1}\\
J_{P_{c2}}(x) &=& \frac{1}{\sqrt{3}} \, \int d^4y \, \Phi_{P_{c2}}(y^2) \
\gamma_\mu \gamma_5 H_c(x+\omega_D^*y) \bar D^{*\mu}(x-\omega_{H_c} y) \,,
\label{Curr_Pc2} \\
J_{P_{c3}}^\mu(x) &=& \int d^4y \, \Phi_{P_{c3}}(y^2) \ 
H_c(x+\omega_D^*y) \bar D^{*\mu}(x-\omega_{H_c}y) \,. 
\label{Curr_Pc3}  
\en 
$H_c$ denotes a single-charm baryon, 
$\Phi_{P_{ci}}(y^2)$ is a phenomenological correlation function 
describing the distribution of $H_c \bar D(\bar D^{*})$ in the pentaquark state $P_{ci}$, 
$\omega_{H_c} = M_{H_c}/(M_{H_c} + M_{D^{(*)}})$ 
and $\omega_{D^{(*)}} = M_{D^{(*)}}/(M_{H_c} + M_{D^{(*)}})$ 
are the mass fractions 
of the constituent hadrons with $\omega_{H_c} + \omega_{D^{(*)}} = 1$. 
Here we include an overall factor $1/\sqrt{3}$ for the case of the interpolating 
current of pentaquark $P_{c2}$ in order to have the same results 
for the couplings of all pentaquarks in heavy quark limit (HQL), 
i.e. in the limit when heavy quark mass goes to infinity. 
To generate ultraviolet-finite Feynman diagrams,
the Fourier transform of the correlation function $\Phi_{P_{ci}}(y^2)$
should vanish sufficiently fast in the ultraviolet region of 
the Euclidean space.
We use the Gaussian form for the correlation function
$\tilde\Phi_{P_{ci}}(p^2_E) \doteq \exp(-p^2_E /\Lambda_{P_{ci}}^2)$, 
where $p_E$ is the Euclidean Jacobi momentum and $\Lambda_{P_{ci}}$
is a free size parameter. 

The coupling $g_{P_{c}}$ is determined from the compositeness
condition (see Refs.~\cite{Salam:1962ap}-\cite{Efimov:1993ei}
and~\cite{Faessler:2007gv}-\cite{Dong:2017gaw}) 
\eq\label{ZPci}
Z_{P_{c1/c2}} &=& 1 - \Sigma_{P_{c1/c2}}^{\prime}(M_{P_{c1/c2}})\equiv 0 \,,\\
Z_{P_{c3}}    &=& 1 - \Sigma_{P_{c3}}^{T \prime}(M_{P_{c3}})    \equiv 0 \,,
\en
where $\Sigma_{P_{c1/c2}}^{\prime}$ and $\Sigma_{P_{c3}}^{T \prime}$ 
are the derivatives of the full and the transverse part 
of the mass operator of the $P_{c1/c2}$ and $P_{c3}$ states, respectively.  
Here we have
\eq
\Sigma_{P_{c3}}^{\mu\nu}(p) =
       g_{\perp}^{\mu\nu}(p) \, \Sigma_{P_{c3}}^T(p) 
+ \frac{p^\mu p^\nu}{p^2} \Sigma_{P_{c3}}^L(p),
\en
where $g^{\mu\nu}_{\perp}(p)=g^{\mu\nu}-p^\mu p^\nu /p^2$. 
The generic Feynman diagram representing the mass operator  
$\Sigma_{P_c}$, which is generated by loop containing the 
$(H_c \bar D)$ or $(H_c\bar D^*)$ constituents, is shown in Fig.~1 
(left panel). Note that the compositeness condition gives a relation 
between the coupling constant $g_{P_{c}}$ and the mass $m_{P_{c}}$.  

The extension to the bottom sector is straightforward. Now we have to construct 
phenomenological Lagrangians describing the coupling of the hidden bottom 
pentaquark to its constituents by the following replacements  
in Eqs.~(\ref{Lagr_Pc})-(\ref{Curr_Pc3}):  
\eq 
P_{ci} \to P_{bi}\,, \quad D(D^*) \to B(B^*)\,, 
\quad \{\Sigma_c,\Xi_c',\Omega_c\} \to \{\Sigma_b,\Xi_b',\Omega_b\}\,.
\en 
The corresponding Feynman diagram for the mass operator
$\Sigma_{P_b}$, which is generated by loop containing the
$(H_b B)$ or $(H_b B^*)$ constituents, is shown in Fig.~1
(right panel).

The classification of the hidden bottom pentaquarks composed of 
single bottom baryons and $B(B^*)$ mesons is presented 
in Table~\ref{tab:hadronsb}. For an estimate of their masses we use 
some typical values close to the corresponding thresholds. 

\begin{figure}
\begin{center}
\epsfig{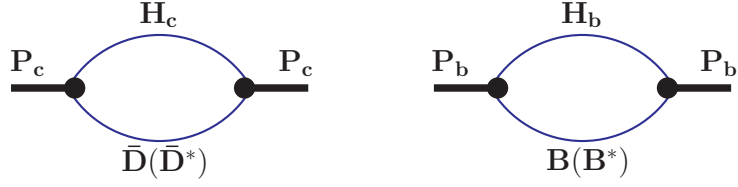}
\caption{Diagrams representing the mass operator of the hidden charm $P_{c}$ 
and bottom $P_{b}$ pentaquarks.} 
\label{fig:Mass_M}
\end{center}
\end{figure}

The expressions for the mass operators $\Sigma_{P_Q}$ ($Q=c,b$) 
are given by 
\eq
\Sigma_{P_{Q1}}(p) &=& 
g_{P_{Q1}}^2 \,
\int \frac{d^4k}{(2\pi)^4 i} \tilde\Phi^2_{P_{Q1}}\Big(-(k+p\omega_P)^2\Big) 
\, S_{H_Q}(k+p) \, S_P(k) \,, \\
\Sigma_{P_{Q2}}(p) &=& 
g_{P_{Q2}}^2 \,
\int \frac{d^4k}{(2\pi)^4 i} \tilde\Phi^2_{P_{Q2}}\Big(-(k+p\omega_V)^2
\Big) \,
\gamma_\mu \gamma_5 \, S_{H_Q}(k+p) \, \gamma_\nu \gamma_5 \, 
S_V^{\mu\nu}(k) \,, \\
\Sigma_{P_{Q3}}^{T}(p) &=& - \frac{1}{3} g_{\perp, \mu\nu}(p) \, 
g_{P_{Q3}}^2 \,
\int \frac{d^4k}{(2\pi)^4 i} \tilde\Phi^2\Big(-(k+p\omega_V)^2\Big) \,
S_{H_Q}(k+p) \, S_V^{\mu\nu}(k) \,, 
\en
where 
\eq 
S_{H_Q}(k) = \frac{1}{M_{H_Q} - \not\! k} \,, \quad 
S_P(k) = \frac{1}{M_P^2 - k^2} \,, \quad 
S_V^{\mu\nu}(k) = - \frac{g_{\perp}^{\mu\nu}(k)}{M_V^2 - k^2} 
\en 
are the propagators of 
the spin-$\frac{1}{2}$ baryon $H_Q$, $P = D (B)$, and $V = D^* (B^*)$ mesons, 
respectively.  
 
\begin{figure}[hb]
\begin{center}
\epsfig{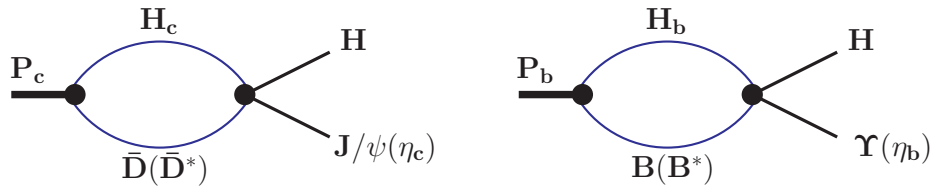}
\caption{Diagrams describing strong two-body decays 
$P_c \to H J/\psi \, (\eta_c)$ and $P_b \to H \Upsilon \, (\eta_b)$, 
where $H$ represents a light baryon octet state} 
\label{fig:decay}
\end{center}
\end{figure}

The leading diagrams contributing to the strong two-body decays 
$P_c \to H J/\psi (\eta_c)$ and $P_b \to H \Upsilon (\eta_b)$, 
are shown in Fig.~\ref{fig:decay}. Here $H$ is a light baryon octet state 
corresponding to the light quark flavor content of the decaying pentaquark. 
The two-body decays $P_c \to H J/\psi (\eta_c)$ and $P_b \to H \Upsilon (\eta_b)$
proceed via the hadronic loops $H \bar D (\bar D^*)$ and $H B (B^*)$,  
composed of the constituents contained in the hidden heavy pentaquark 
state. To evaluate the diagrams in Fig.~\ref{fig:decay} we need to set up an 
interaction Lagrangian, which includes the 
coupling of the pentaquark constituents to the final hadrons 
($H + M$ pair, where $M = J/\psi, \eta_c, \Upsilon, \eta_b$). 
We need to specify three types of interaction Lagrangians: 
(1) ${\cal L}_{H_iH_fPV}$   -- 
the coupling of heavy-light $H_i = H_c, H_b$ and light $H_f = H$ baryons to 
pseudoscalar $P = D, B, \eta_c, \eta_b$ and vector $V = D^*, B^*, J/\psi, 
\Upsilon$ mesons,   
(2) ${\cal L}_{H_iH_fP_1P_2}$ -- 
the coupling of $H_i$ and $H_f$ to two pseudoscalar mesons
$P_1 = D, B$ and $P_2 = \eta_c, \eta_b$, 
(3) ${\cal L}_{H_iH_fVV}$ -- 
the coupling of $H_i$ and $H_f$ to two vector mesons 
$V_1 = D^*, B^*$ and $V_2 = J/\psi, \Upsilon$. 

The $SU(4)$ symmetric form of the first phenomenological Lagrangian, 
${\cal L}_{H_iH_fPV}$ was originally derived in Refs.~\cite{Okubo:1975sc} 
and extensively employed in our formalism 
in Refs.~\cite{Dong:2009tg,Dong:2011tg}. Here we extend
this Lagrangian to five flavors (we only display the part of the
Lagrangian which contributes to current processes) with 
\eq 
{\cal L} = g \bar H^{kmn} i \gamma^\mu \gamma^5 \, [V_\mu,P]^{kl} \, H^{lnm} 
\, + \, {\rm H.c.}\, .
\en
The commutator of vector and pseudoscalar mesons is given by
$[V_\mu,P] = V_\mu P - P V_\mu$, $H^{kmn}$ is the baryon field and 
$(k,m,n,l = 1,\ldots,5)$ is the set of $SU(5)$ flavor indices.
The effective coupling
\eq\label{g1} 
g = - \frac{g_V}{f_P \sqrt{2}} \,
\en 
was already fixed in Refs.~\cite{Dong:2009tg}, where 
$g_V = g_{\rho NN} = 4.8$~\cite{Gutsche:2016jap} is the strong $\rho NN$ 
coupling constant and $f_P$ is the pseudoscalar decay constant, which 
for the charm and bottom sectors is identified with 
$f_{\eta_c}$ and $f_{\eta_b}$. The leptonic decay constants $f_{\eta_c}$ 
and $f_{\eta_b}$ have been evaluated in Lattice QCD~\cite{Chiu:2007km}: 
$f_{\eta_c} = 438 \pm 5 \pm 6$ MeV and 
$f_{\eta_b} = 801 \pm 7 \pm 5$ MeV and in several phenomenological 
approaches (see, e.g., Refs.~\cite{Hwang:1997ie,Gutsche:2016jap}).  
In particular, in Ref.~\cite{Hwang:1997ie} the leptonic decay 
constants of heavy quarkonia have been predicted using  
the Royen-Weisskopf formula: $f_{\eta_c} = 420 \pm 52$ MeV and 
$f_{\eta_b} = 705 \pm 27$ MeV. In Ref.~\cite{Gutsche:2016jap} these couplings 
have been estimated using the covariant confined quark model: 
$f_{\eta_c} = 427$ MeV and $f_{\eta_b} = 772$ MeV. In our calculations 
we will use averaged values of the three sources: 
$f_{\eta_c} = 430$ MeV and $f_{\eta_b} = 750$ MeV. 
        
The 
expression of the physical meson and baryon states in terms 
of the tensors $V^{mn}$, $P^{mn}$, and $H^{mnk}$ are discussed in detail 
in Refs.~\cite{Okubo:1975sc,Dong:2009tg,Dong:2011tg}. Here we present 
some examples:  
\eq 
& &p = H^{112} = - 2 H^{121} = - 2 H^{211}\,, \quad  
   n = H^{221} = - 2 H^{212} = - 2 H^{122}\,, \nonumber\\
& &\Sigma_c^{++} = H^{114} = - 2 H^{141} = - 2 H^{411}\,, \quad  
   \Sigma_b^{+}  = H^{115} = - 2 H^{151} = - 2 H^{511}\,, \nonumber\\ 
& &J/\psi = V^{44}, \quad  \Upsilon = V^{55}, \quad  
   \eta_c = P^{44}, \quad  \eta_b = P^{55}, \\ 
& &\bar D^0 = P^{14}, \quad  D^- = P^{24}, \quad 
   \bar D^{*0} = V^{14}, \quad  D^{*-} = V^{24}, \nonumber\\ 
& &B^+ = P^{51}, \quad  B^0 = P^{52}, \quad 
   B^{*+} = V^{51}, \quad  B^{*0} = V^{52}\,. 
\nonumber
\en 
Part of this Lagrangian 
involving the coupling of the light-heavy light baryon pair 
to $DJ/\psi$, $D^*\eta_c$, $B\Upsilon$, and $B^*\eta_b$ pairs  
needed for our calculation reads 
\eq 
{\cal L} = g \, 
\biggl[      M_c^{- \mu} \, J^{c+}_{\mu} 
\, + \, \bar M_c^{0 \mu} \, J^{c0}_{\mu} 
\, + \,      M_b^{+ \mu} \, J^{b-}_{\mu}
\, + \,      M_b^{0 \mu} \, J^{b0}_{\mu}
\biggr]
\, + \,     {\rm H.c.}\,. 
\en 
Here we introduce the notations 
\eq 
M_c^{- \mu} &=& \eta_c \, D^{*-}_\mu \, - \,  J/\psi_\mu \, D^- \,, 
\nonumber\\
\bar M_c^{0 \mu} &=& \eta_c \, \bar D^{*0}_\mu \, - \,  
J/\psi_\mu \, \bar D^0 \,, 
\nonumber\\
M_b^{+ \mu} &=& \eta_b \, B^{*+}_\mu \, - \, \Upsilon_\mu \, B^+ \,, 
\nonumber\\
M_b^{0 \mu} &=& \eta_b \, B^{*0}_\mu \, - \, \Upsilon_\mu \, B^0 \,, 
\en 
and 
$J^{c+/b-}_{\mu}$, $J^{c0/b0}_{\mu}$ are the charged and neutral 
axial-vector currents composed of charm (bottom) and light baryon as 
\eq 
J^{c+}_{\mu} &=& \frac{1}{4} \biggl[ \bar p \gamma_\mu \gamma_5 \Sigma^{++}_c 
  - \frac{1}{\sqrt{2}} \, \bar n \gamma_\mu \gamma_5 \Sigma^{+}_c 
  - \frac{1}{\sqrt{2}} \, \bar \Sigma^- \gamma_\mu \gamma_5 \Xi^{\prime 0}_c 
  - \frac{1}{2} \, \bar \Sigma^0  \gamma_\mu \gamma_5 \Xi^{\prime +}_c 
  + \bar \Xi^-  \gamma_\mu \gamma_5 \Omega_c^0 
  - \frac{\sqrt{3}}{2} \bar \Lambda^0 \gamma_\mu \gamma_5 \Xi^{'+}_c \biggr]
\,, 
\label{JVpcurrent}\\
J^{c0}_{\mu} &=& \frac{1}{4} \biggl[ \bar n \gamma_\mu \gamma_5 \Sigma^{0}_c 
  - \frac{1}{\sqrt{2}} \, \bar p \gamma_\mu \gamma_5 \Sigma^{+}_c 
  - \frac{1}{\sqrt{2}} \, \bar \Sigma^+ \gamma_\mu \gamma_5 \Xi^{\prime +}_c 
  - \frac{1}{2} \, \bar \Sigma^0  \gamma_\mu \gamma_5 \Xi^{\prime 0}_c 
  + \bar \Xi^0 \gamma_\mu \gamma_5 \Omega_c^0 
  + \frac{\sqrt{3}}{2} \bar \Lambda^0 \gamma_\mu \gamma_5 \Xi^{'0}_c \biggr]
\,,
\label{JV0current}\\ 
J^{b0}_{\mu} &=& \frac{1}{4} \biggl[ \bar p \gamma_\mu \gamma_5 \Sigma^{+}_b 
  - \frac{1}{\sqrt{2}} \, \bar n \gamma_\mu \gamma_5 \Sigma^{0}_b 
  - \frac{1}{\sqrt{2}} \, \bar \Sigma^- \gamma_\mu \gamma_5 \Xi^{\prime -}_b 
  - \frac{1}{2} \, \bar \Sigma^0  \gamma_\mu \gamma_5 \Xi^{\prime 0}_b 
  + \bar \Xi^-  \gamma_\mu \gamma_5 \Omega_b^-
  - \frac{\sqrt{3}}{2} \bar \Lambda^0 \gamma_\mu \gamma_5 \Xi^{'0}_b \biggr]
\,, 
\label{JVbpcurrent}\\
J^{b-}_{\mu} &=& \frac{1}{4} \biggl[ \bar n \gamma_\mu \gamma_5 \Sigma^{-}_b
  - \frac{1}{\sqrt{2}} \, \bar p \gamma_\mu \gamma_5 \Sigma^{0}_b 
  - \frac{1}{\sqrt{2}} \, \bar \Sigma^+ \gamma_\mu \gamma_5 \Xi^{\prime 0}_b 
  - \frac{1}{2} \, \bar \Sigma^0  \gamma_\mu \gamma_5 \Xi^{\prime -}_b
  + \bar \Xi^0 \gamma_\mu \gamma_5 \Omega_b^-  
  + \frac{\sqrt{3}}{2} \bar \Lambda^0 \gamma_\mu \gamma_5 \Xi^{'-}_b \biggr]
\,.
\label{JVb0current}
\en  
Using Eqs.~(\ref{JVpcurrent}) and~(\ref{JV0current}) we derive the couplings 
$g_{H_cHDJ/\psi} = - g_{H_cH\eta_cD^*} = c_{HH_c} \, g$ 
and 
$g_{H_bHB\Upsilon} = - g_{H_bH\eta_cB^*} = c_{HH_b} \, g$,  
where $c_{HH_c}$ and $c_{HH_b}$ are flavor factors shown in 
Table~\ref{tab:BBpDJcouplings}. Note that above results 
for the couplings are consistent with recent predictions derived on the basis 
of heavy-flavor and spin symmetry 
in Refs.~\cite{Voloshin:2019aut,Sakai:2019qph}. We also derive 
the relations between couplings involving charm and bottom hadrons: 
\eq 
\frac{g_{H_cHDJ/\psi}}{g_{H_bHDJ/\psi}} = 
\frac{g_{H_cH\eta_cD^*}}{g_{H_bH\eta_bD^*}} = 
\frac{c_{HH_c}}{c_{HH_b}} \,. 
\en

In the next step we introduce the couplings of two pseudoscalar and two vector mesons 
to a baryon pair implementing the consequences of heavy-flavor and spin symmetry. 
We are therefore consistent with the results of Refs.~\cite{Voloshin:2019aut,Sakai:2019qph}. 
In this vein we derive the interaction Lagrangian required for the description of 
the decays of hidden flavor pentaquarks to $J/\psi (\eta_c) + H$ and $\Upsilon (\eta_b) + H$ 
pairs. The relevant Lagrangian reads: 
\eq
{\cal L}_{\rm int} &=& g \, c_{HH_c} \, \bar H \, 
\biggl\{ \, \Big[5 \, J/\psi^{\,\mu} 
\, + \, \eta_c \, i \gamma^\mu \gamma^5\Big] \, \bar D^*_\mu 
\, + \, \Big[ J/\psi_\mu \, i \gamma^\mu \gamma^5 
\, - \, 3 \, \eta_c \Big] \, \bar D \, 
\biggr\} \, H_c  \nonumber\\
&+& g \, c_{HH_b} \, \bar H \, 
\biggl\{ \, \Big[5 \, \Upsilon^\mu 
\ \, \, \, \, \,  + \, \eta_b \, i \gamma^\mu \gamma^5\Big] \, B^*_\mu 
\, + \, \, \Big[ \Upsilon_\mu \, i \gamma^\mu \gamma^5 \, \ \, \,  
\, - \, 3 \, \eta_b \Big] \, B \,  
\biggr\} \, H_b + {\rm H.c.}  
\en 
Note that again the coupling $g$ is fixed [see Eq.~(\ref{g1})] 
and expressed in terms of well-known couplings/parameters. 
The $P_{Q3}$ pentaquark cannot decay into $\eta_Q H$ in an 
$S$-wave as was stressed in Ref.~\cite{Voloshin:2019aut}, 
while it can proceed in a $D$ wave. We introduce an additional 
$D$-wave coupling and specify it by introducing an additional parameter $\beta$ 
(for convenience we scale it by a factor $M_{P_{Q3}}$):  
\eq 
{\cal L}_{\rm int}^{\rm D-wave} = \frac{\beta}{M_{P_{c3}}} \, 
g \, c_{HH_c} \, \bar H \, \partial^\mu\eta_c \, \gamma^5 \, \bar D^*_\mu \, H_c 
\,+\, \frac{\beta}{M_{P_{b3}}} \, g \, c_{HH_b} \, \bar H \, \partial^\mu\eta_b 
\, \gamma^5 \, B^*_\mu \, H_b  \,+\, {\rm H.c.}
\en 
Such an additional coupling leads to a suppression of the decay rates for 
$P_{Q3} \to \eta_Q H$  by a factor $1/m_Q$ in comparison 
to the modes $P_{Q1} \to \eta_Q H$ and $P_{Q2} \to \eta_Q H$. 
It also  gives a contribution to the matrix element of the 
$P_{Q2} \to \eta_Q H$ transition at next-to-leading order 
in the heavy quark mass expansion, and, therefore, will be suppressed 
in comparison with the $S$-wave. 
 
\begin{table}[htb]
\begin{center}
\caption{Flavor factors $c_{HH_c}$ and $c_{HH_b}$}

\vspace*{.1cm}

\def\arraystretch{1.2}
\begin{tabular}{c|c||c|c}
\hline
\ \ Flavor structure $HH_c$ \ \
& \ \ $c_{HH_c}$ \ \ &
\ \ Flavor structure $HH_b$ \ \
& \ \ $c_{HH_b}$ \ \ \\
\hline
  $p\Sigma_c^{++}$ & $\frac{1}{4}$
& $p\Sigma_b^{+}$  & $\frac{1}{4}$ \\
\hline
  $n\Sigma_c^{0}$  & $\frac{1}{4}$
& $n\Sigma_b^{-}$  & $\frac{1}{4}$ \\
\hline
  $p\Sigma_c^{+}$  & $- \frac{1}{4 \sqrt{2}}$
& $p\Sigma_b^{0}$  & $- \frac{1}{4 \sqrt{2}}$ \\
\hline
  $n\Sigma_c^{+}$  & $- \frac{1}{4 \sqrt{2}}$
& $n\Sigma_b^{0}$  & $- \frac{1}{4 \sqrt{2}}$ \\
\hline
  $\Sigma^-\Xi_c^{'0}$  & $- \frac{1}{4 \sqrt{2}}$
& $\Sigma^-\Xi_b^{'-}$  & $- \frac{1}{4 \sqrt{2}}$ \\
\hline
  $\Sigma^+\Xi_c^{'+}$  & $- \frac{1}{4 \sqrt{2}}$
& $\Sigma^+\Xi_b^{'0}$  & $- \frac{1}{4 \sqrt{2}}$ \\
\hline
  $\Sigma^0\Xi_c^{'+}$  & $- \frac{1}{8}$
& $\Sigma^0\Xi_b^{'0}$  & $- \frac{1}{8}$ \\
\hline
  $\Sigma^0\Xi_c^{'0}$  & $- \frac{1}{8}$
& $\Sigma^0\Xi_b^{'-}$  & $- \frac{1}{8}$ \\
\hline
  $\Lambda^0\Xi_c^{'+}$  & $- \frac{\sqrt{6}}{8}$
& $\Lambda^0\Xi_b^{'0}$  & $- \frac{\sqrt{6}}{8}$ \\
\hline
  $\Lambda^0\Xi_c^{'0}$  & $  \frac{\sqrt{6}}{8}$
& $\Lambda^0\Xi_b^{'-}$  & $  \frac{\sqrt{6}}{8}$ \\
\hline
  $\Xi^- \Omega_c^0$    & $\frac{1}{4}$
& $\Xi^- \Omega_b^-$    & $\frac{1}{4}$   \\
\hline
  $\Xi^0 \Omega_c^0$    & $\frac{1}{4}$
& $\Xi^0 \Omega_b^-$    & $\frac{1}{4}$   \\
\hline
\end{tabular}
\label{tab:BBpDJcouplings}
\end{center}

\begin{center}
\caption{Flavor factors $d_H$}

\vspace*{.1cm}

\def\arraystretch{1.5}
\begin{tabular}{c|c}
\hline
\ \ Light baryon $H$ \ \
& \ \ \ \ \ \ $d_{H}$ \ \ \ \ \ \ \\
\hline
  $p$ &   $\frac{3}{4 \sqrt{6}}$ \\
  $n$ & - $\frac{3}{4 \sqrt{6}}$ \\
  $\Sigma^{+}$ & - $\frac{1}{4 \sqrt{2}}$ \\
  $\Sigma^{0}$ & - $\frac{1}{4 \sqrt{2}}$ \\
  $\Sigma^{-}$ & - $\frac{1}{4 \sqrt{2}}$ \\
  $\Lambda^{0}$ & - $\frac{\sqrt{3}}{4}$ \\
  $\Xi^{0}$ & $\frac{1}{4}$ \\
  $\Xi^{-}$ & $\frac{1}{4}$ \\
\hline
\end{tabular}
\label{tab:dBBpDJcouplings}
\end{center}
\end{table}

In the calculation of the two-body decays $P_{ci} \to H + M$ 
($M = V,P$ and $V = J/\psi, \Upsilon$; \ $P = \eta_c, \eta_b$)  
of pentaquark $P_{ci}$ with spin $S_{P_{ci}}$ we use the rest frame of 
the pentaquark with the final baryon moving in the positive $z$-direction. 
The 4-momenta of pentaquark ($p_1$), final baryon ($p_2$), 
and meson ($p_3$) are specified as 
\eq 
p_1 = (M_{P_{ci}}, {\bf 0}\,)\,, \quad 
p_2 = (E_{H},0,0, |{\bf p_2}\,|)\,, \quad 
q = p_1 - p_2 = (E_{M},0,0,-|{\bf p_2}\,|)\,. 
\en 
We will also use the following notations: 
$M_\pm = M_{P_{ci}} \pm M_M$\,, 
$Q_\pm = M_\pm^2 - M_{M}^2$  
and 
$\lambda(x,y,z) = x^2 + y^2 + z^2 - 2xy - 2xz - 2yz$ 
is the K\"allen kinematical triangle function. 
Energy values and three-momenta of 
the decay products are defined as 
\eq 
E_H &=& \frac{M_{P_{ci}}^2 + M_H^2 - M_M^2}{2 M_{P_{ci}}}\,, 
\nonumber\\
E_M &=& M_{P_{ci}} - E_H = 
\frac{M_{P_{ci}}^2 - M_H^2 + M_M^2}{2 M_{P_{ci}}}\,, 
\\
|{\bf p_2}\,| &=& 
\frac{\lambda^{1/2}(M_{P_{ci}}^2,M_H^2,M_M^2)}{2 M_{P_{ci}}}  
= \frac{\sqrt{Q_+Q_-}}{2 M_{P_{ci}}} \;.  
\en 
Due to Lorentz covariance and because of the transversity condition 
$q_\mu \, \epsilon_V^\mu  = 0$ 
for the polarization vector of the $V = J/\psi, \Upsilon$ mesons,  
the matrix elements of the $P_{ci} \to H + V$ decay processes are 
in general expressed in terms of 
two form factors ($F_i^V, i=1,2$) for the 
$\frac{1}{2}^- \to \frac{1}{2}^+ + 1^-$ transitions 
and three ($F_i^V, i=1,2,3$) for the 
$\frac{3}{2}^- \to \frac{1}{2}^+ + 1^-$ transitions 
(see details in Refs.~\cite{Faessler:2009xn,Gutsche:2017wag}): 

\noindent
transition $\frac{1}{2}^- \to \frac{1}{2}^+ + 1^-$\,:
\eq
M_{\rm inv}\biggl(\frac{1}{2}^- \to \frac{1}{2}^+ + 1^-\biggr) 
= \bar u(p_2,s_2)
\Big[ \gamma_\mu F_1^V(M_V^2)
    - i \sigma_{\mu\nu} \frac{q^\nu}{M_{P_{ci}}} F_2^V(M_V^2) 
\Big]\gamma_5
u(p_1,s_1) \, \epsilon^\mu_V(q) \, ,
\en

\noindent
transition $\frac{3}{2}^- \to \frac{1}{2}^+ + 1^-$\,:

\eq
M_{\rm inv}\biggl(\frac{3}{2}^- \to \frac{1}{2}^+ + 1^-\biggr) 
= \bar u(p_2,s_2)
\Big[ g_{\alpha\mu} F_1^V(M_V^2) 
    + \gamma_\mu \frac{p_{2\alpha}}{M_{P_{ci}}} F_2^V(M_V^2) 
+ \frac{p_{2\alpha} p_{2\mu}}{M_{P_{ci}}} F_3^V(M_V^2) 
\Big]  u^\alpha(p_1,s_1) \, \epsilon^\mu_V(q) \,, 
\en
where
$\sigma_{\mu\nu} = (i/2) (\gamma_\mu \gamma_\nu - \gamma_\nu \gamma_\mu)$
and all $\gamma$ matrices are defined as in the Bjorken-Drell convention. 
In our approach only one form factor $F_1^V$ contributes
to the transitions $P_{c1} \to H + V$ 
and $P_{c3} \to H + V$, while the others vanish. 

For the decay modes involving $P = \eta_c, \eta_b$ in the final state 
the matrix elements are expressed in terms of a single 
pseudoscalar form factor $F^P(M_P^2)$: 

\noindent
transition $\frac{1}{2}^- \to \frac{1}{2}^+ + 0^-$\,:
\eq 
M_{\rm inv}\biggl(\frac{1}{2}^- \to \frac{1}{2}^+ + 0^-\biggr) 
= \bar u(p_2,s_2) \, F^P(M_P^2) \, u(p_1,s_1) \, , 
\en 

\noindent
transition $\frac{3}{2}^- \to \frac{1}{2}^+ + 0^-$\,:
\eq 
M_{\rm inv}\biggl(\frac{3}{2}^- \to \frac{1}{2}^+ + 0^-\biggr) 
= \bar u(p_2,s_2) \, F^P(M_P^2) \, 
i \gamma^5 \frac{q_\alpha}{M_{P_{Q3}}} \, u^\alpha(p_1,s_1) 
\,. \nonumber
\en

It is convenient to express the decay widths of the two-body decays 
$P_{ci} \to H + V$ in terms of the helicity amplitudes 
$H_{\lambda_2\lambda_V}$~\cite{Faessler:2009xn,Gutsche:2017wag}, 
where $\lambda_V = \pm 1, 0$ and 
$\lambda_2 = \pm 1/2, \pm 3/2$ are  the helicity components of the $V$ 
mesons and the final baryon $H$, respectively. For our kinematics 
helicity conservation reads: $\lambda_1 = \lambda_2 - \lambda_V$, 
where $\lambda_1$ is the helicity of decaying pentaquark. 
The helicity amplitudes are related to the sets of the previously introduced 
relativistic form factors $F_i^V$ as~\cite{Faessler:2009xn,Gutsche:2017wag}: 

\noindent
transition $\frac{1}{2}^-\to \frac{1}{2}^+ + 1^-$\,:
$H^V_{-\lambda_2,-\lambda_V} = - H^V_{\lambda_2,\lambda_V}$

\eq 
H_{\frac{1}{2} 0}^V &=& \sqrt{\frac{Q_+}{M_V^2}} \,
\Big( F_1^V M_- - F_2^V \frac{M_V^2}{M_{P_{ci}}} \Big)\,, \nonumber\\
H_{\frac{1}{2} 1}^V &=& \sqrt{2Q_+}
\Big( - F_1^V + F_2^V \frac{M_-}{M_{P_{ci}}} \Big)  \, ,
\en

transition $\frac{1}{2}^-\to \frac{1}{2}^+ + 0^-$\,:
$H^P_{-\frac{1}{2}t} = - H^P_{\frac{1}{2}t}$

\eq 
H_{\frac{1}{2} t}^P &=& \sqrt{Q_+} \, F^P \;,
\en

transition $\frac{3}{2}^-\to \frac{1}{2}^+ + 1^-$\,:
$H^V_{-\lambda_2,-\lambda_V} = H^V_{\lambda_2,\lambda_V}$

\eq
H_{\frac{1}{2} 0}^V &=&  \sqrt{\frac{2 Q_+}{3}} \, 
\, \frac{M_+M_- + M_V^2}{2M_{P_{c3}} M_V} \, F_1^V\,, 
\nonumber\\
H_{\frac{1}{2} 1}^V &=& 
\sqrt{\frac{Q_+}{3}} \, F_1^V 
\qquad
H_{\frac{3}{2} 1}^V \, = \, \sqrt{Q_+} F_1^V\,,
\en
we omit the contribution of $F_2^V$ and $F_3^V$ since for the present calculation
these form factors vanish;

transition $\frac{3}{2}^-\to \frac{1}{2}^+ + 0^-$\,:
$H^P_{-\frac{1}{2}t} = H^P_{\frac{1}{2}t}$

\eq 
H_{\frac{1}{2} t}^P &=& \sqrt{\frac{Q_+}{6}} \, \, 
\frac{Q_-}{M_{P_{c3}}^2} \, F^P\,.
\en
In the case of the $\frac{3}{2}^-\to \frac{1}{2}^+ + 0^-$ transition
the helicity amplitude $H_{\frac{1}{2} t}^P$ 
has an additional $1/m_Q$ suppression factor in comparison to the others.
As already explained, the reason lies in the $D$-wave coupling of this 
transition.

The decay width of the two-body transition $P_{ci} \to H + V (P)$ 
is calculated according to the 
formulas~\cite{Faessler:2009xn,Gutsche:2017wag}:  
\eq 
\Gamma(P_{ci} \to H + V)
&=& \frac{1}{8 \pi (2 S_{P_{ci}} + 1)} 
\, \frac{|{\bf p_2}\,|}{M_{P_{ci}}^2} \, {\cal H}_{P_{ci}H}^V\,, 
\quad {\cal H}_{P_{ci}H}^V \, = \,
\sum_{\lambda_2,\lambda_V}|H_{\lambda_2,\lambda_{J/\psi}}^V|^2\,, 
\nonumber\\
\Gamma(P_{ci} \to H + P) 
&=& \frac{1}{8 \pi (2 S_{P_{ci}} + 1)} 
\, \frac{|{\bf p_2}\,|}{M_{P_{ci}}^2} \, {\cal H}_{P_{ci}H}^P\,, 
\quad 
{\cal H}_{P_{ci}H}^P \, = \, \sum_{\lambda_2} \, |H_{\lambda_2, t}^P|^2 \,, 
\en 
where 
$S_{P_{c1}} = S_{P_{c2}} = \frac{1}{2}$ and 
$S_{P_{c3}} = \frac{3}{2}$ are the spins of the pentaquarks, and 
${\cal H}_{P_{ci}H}^{V,P}$ is the sum of the corresponding squared helicity 
amplitudes with: 
\eq 
& &{\cal H}_{P_{c1/c2}H}^V = 
|H_{\frac{1}{2} 0}^V|^2 + |H_{-\frac{1}{2}  0}^V|^2 + 
|H_{\frac{1}{2} 1}^V|^2 + |H_{-\frac{1}{2} -1}^V|^2  \,, 
\nonumber\\
& &{\cal H}_{P_{c3}H}^V = 
|H_{\frac{1}{2} 0}^V|^2 + |H_{-\frac{1}{2}  0}^V|^2 + 
|H_{\frac{1}{2} 1}^V|^2 + |H_{-\frac{1}{2} -1}^V|^2 + 
|H_{\frac{3}{2} 1}^V|^2 + |H_{-\frac{3}{2} -1}^V|^2 \,,
\nonumber\\
& &{\cal H}_{P_{ci}H}^P = 
|H_{\frac{1}{2} t}^P|^2 + |H_{-\frac{1}{2}  t}^P|^2 \,. 
\en 

\section{Results}

We proceed with our calculation in the heavy quark limit (HQL) expanding 
the masses of the heavy hadrons around the corresponding heavy quark masses $m_Q$, $Q=c,b$: 
\eq 
& &M_{P_{Qi}} = 2 m_Q  + {\cal O}(1)\,, \quad 
   M_{J/\psi,\eta_c}  = 2 m_c  + {\cal O}(1)\,, \quad 
   M_{\Upsilon,\eta_b}  = 2 m_b  + {\cal O}(1)\,, \nonumber\\
& &M_{D,D^*} = m_c + {\cal O}(1)\,, \quad 
   M_{B,B^*} = m_b + {\cal O}(1)\,, \\ 
& &M_{H_Q} = m_Q + {\cal O}(1)\,, \quad 
   M_{H}   = {\cal }O(1)\,.
\en 
We also introduce the heavy-flavor independent 
quantity $R$, the difference between the sum of the masses of the 
constituents and the pentaquark mass, with:  
\eq 
R = M_{H_Q} + M_{M_Q} - M_{P_{Qi}} 
\en
which at leading order is considered to be universal for all modes. 

In the HQL the results become rather transparent. In particular, 
the coupling constants of pentaquarks to the constituents scale 
as $\sqrt{m_Q}$ and are given by 
\eq 
g_{P_{Q1}} = g_{P_{Q2}} = g_{P_{Q3}} =  
4 \pi \, \sqrt{\frac{2m_Q}{\Lambda}} \, [I(r)]^{-1/2}\,,
\en  
where $I(r)$, $r=R/\Lambda$ is the structure integral 
\eq 
I(r) = 
\int\limits_0^\infty d\alpha_1 
\int\limits_0^\infty d\alpha_2 
(\alpha_1 + \alpha_2) \, \exp\biggl[-\frac{1}{2} 
(\alpha_1-\alpha_2)^2-\frac{1}{2} (\alpha_1+\alpha_2) r \biggr] \,. 
\en 
The couplings for all three types of pentaquarks are degenerate 
in the HQL. For these reason we introduced an additional factor 
$1/\sqrt{3}$ in the interpolating current of the pentaquark $P_{Q2}$ 
[see Eq.~(\ref{Curr_Pc2})].  

The transition amplitudes in the HQL are also given in terms of 
a single structure integral 
\eq 
J(r) = \int\limits_0^\infty d\alpha_1 
       \int\limits_0^\infty d\alpha_2 
\, \exp\biggl[-(\alpha_1-\alpha_2)^2
-\frac{1}{2} (\alpha_1+\alpha_2) r \biggr] \,. 
\en 
Therefore, all helicity amplitudes only depend on the function 
$\xi(r) = J(r)/\sqrt{I(r)}$ involving the single parameter $r=R/\Lambda$. 
We find that $\xi(r_1)$ changes only slightly when varying
the size parameter $\Lambda$ in the region 0.5-2 GeV.
In particular, for $R \simeq 10$ MeV and $\Lambda$ ranging 
from 0.5 to 2 GeV the quantity $\xi(r)$ changes from 0.7872 to 0.7905. 

In the HQL results for the squared helicity amplitudes are simply expressed as: 
\eq\label{H_list} 
& &{\cal H}^V_{P_{Q1}} 
= \frac{3}{25} \, {\cal H}^V_{P_{Q2}} 
= \frac{3}{50} \, {\cal H}^V_{P_{Q3}} = 
\frac{1}{3} \, {\cal H}^P_{P_{Q1}} = {\cal H}^P_{P_{Q2}} 
= 12 \ \biggl(\frac{g d_H}{\pi}\biggr)^2 \ \Lambda^4 \ 
\xi^2(r) \ (r + \mu_H)
\,\, \\
& &{\cal H}^P_{P_{Q3}} = \beta^2 \, \frac{M_H}{54 m_Q} 
\, {\cal H}^P_{P_{Q1}} 
\,, 
\en 
where $\mu_H = M_H/\Lambda$ is the mass of the light baryon in the final 
state rescaled by $\Lambda$. The flavor factors $d_H$ depend on the 
flavor of the final baryon and are summarized 
in Table~\ref{tab:dBBpDJcouplings}.

In the HQL the helicity amplitudes ${\cal H}^V_{P_{Qi}}$ and ${\cal H}^P_{P_{Q1/Q2}}$  
scale as ${\cal O}(1/m_Q)$ since the 
heavy quarkonia decay constants, contained in the definition of the coupling $g$,
have the scaling behavior $f_{H_{Q\bar Q}} \sim \sqrt{m_Q}$. 
Therefore, the decay rates of pentaquarks into heavy charmonia and a light baryon 
scale as $1/m_Q^3$ (we take into account that $|{\bf p_2}\,|$ 
behaves as ${\cal O}(1)$ in the HQL).  
All relations between the helicity amplitudes 
(further below it will also be displayed for the decay rates) are consistent 
with the results reported previously in Refs.~\cite{Voloshin:2019aut,Sakai:2019qph}. 
In our approach the $P_{Q3}$ pentaquark decay into 
$\eta_Q H$ is restricted to an orbital $D$-wave, the $S$-wave is forbidden. 
The helicity amplitude for the decay $P_{Q3} \to \eta_Q H$ ($D$-wave) 
contains an arbitrary 
parameter $\beta$, which could be fixed from future experimental results.
However ${\cal H}^P_{P_{Q3}}$ has an additional suppression factor $1/m_Q$ 
in comparison with the other helicity amplitudes listed in Eq.~(\ref{H_list}) 
and the corresponding decay rate scales as $1/m_Q^4$, i.e. it is suppressed 
by a factor $1/m_Q$ relative to the other modes. 

Neglecting the mass differences of $J/\psi-\eta_c$ and $\Upsilon-\eta_b$ 
we finally derive following relations for the pentaquark decay rates: 
\eq 
& &
\frac{\Gamma(P_{c1} \to \eta_c p)}{\Gamma(P_{c1} \to J/\psi p)} = 3\,, 
\quad 
\frac{\Gamma(P_{c2} \to \eta_c p)}{\Gamma(P_{c2} \to J/\psi p)} 
= \frac{3}{25}\,, \quad
\frac{\Gamma(P_{c3} \to \eta_c p)}{\Gamma(P_{c3} \to J/\psi p)} 
= \frac{\beta^2}{300} \, \frac{M_H}{m_c} 
\,, \\
& &\frac{\Gamma(P_{b1} \to \eta_b p)}{\Gamma(P_{b1} \to \Upsilon p)} = 3\,, 
\quad 
\frac{\Gamma(P_{b2} \to \eta_b p)}{\Gamma(P_{b2} \to \Upsilon p)} 
= \frac{3}{25}\,, \quad 
\frac{\Gamma(P_{b3} \to \eta_b p)}{\Gamma(P_{b3} \to \Upsilon p)} 
= \frac{\beta^2}{300} \, \frac{M_H}{m_b}\,, \\
& &\frac{\Gamma(P_{ci} \to \eta_c p)}{\Gamma(P_{bi} \to \eta_b p)} 
= \frac{\Gamma(P_{ci} \to J/\psi p)}{\Gamma(P_{bi} \to \Upsilon p)} 
= \biggl(\frac{m_b}{m_c}\biggr)^3\,, \quad i = 1, 2, 3\,.
\en 
In the following we turn to the discussion of our numerical results. 
We first look at the decay rates of charm nonstrange pentaquarks to the final states 
$J/\psi N$ and $\eta_c N$. The results for the decay rates are expressed in terms 
of the model parameters $\Lambda$ and $\beta$ as 
\eq 
& &\Gamma(P_{c1}^+ \to J/\psi p) = 
4.12 \ {\rm MeV} \ \cdot\, \biggl(\frac{\Lambda}{1 \ \mbox{GeV}}\biggr)^3\,, 
\\
& &\Gamma(P_{c2}^+ \to J/\psi p) = 
39.82 \ {\rm MeV} \ \cdot\, \biggl(\frac{\Lambda}{1 \ \mbox{GeV}}\biggr)^3\,, 
\\
& &\Gamma(P_{c3}^+ \to J/\psi p) = 
80.04 \ {\rm MeV} \ \cdot\,\biggl(\frac{\Lambda}{1 \ \mbox{GeV}}\biggr)^3\,, 
\\
& &\Gamma(P_{c1}^+ \to \eta_c p) = 
14.81 \ {\rm MeV} \ \cdot\, \biggl(\frac{\Lambda}{1 \ \mbox{GeV}}\biggr)^3\,, 
\\
& &\Gamma(P_{c2}^+ \to \eta_c p) =
5.45 \ {\rm MeV} \ \cdot\, \biggl(\frac{\Lambda}{1 \ \mbox{GeV}}\biggr)^3\,,
\\
& &\Gamma(P_{c3}^+ \to \eta_c p) = 
0.13 \ {\rm MeV} \ \cdot\, \beta^2 \,\cdot\, \biggl(\frac{\Lambda}{1 \ \mbox{GeV}}\biggr)^3
\,. 
\en

In Tables~\ref{results:charm} and~\ref{results:bottom} we present the results for other 
decay modes. For many of the indicated states we use typical values for the masses. 
In all results we drop the factor $(\Lambda/1 \ {\rm GeV})^3$. 
 
\begin{table}[hb]
\begin{center}
\caption{Two-body decay rates of hidden charm pentaquarks in MeV  
[results for decay 
rates should be multiplied by factor $(\Lambda/1 \ {\rm GeV})^3$]} 

\vspace*{.1cm}

\def\arraystretch{1}
\begin{tabular}{c|c||c|c}
\hline
   \ \ Mode \ \ 
&  \ \ Decay rate (MeV) \ \ 
&  \ \ Mode \ \ 
&  \ \ Decay rate (MeV) \ \ \\
\hline
 \multicolumn{4}{c}{First Family $P_{c1}$} \\
\cline{1-4}
\hline
$P_{c1}^+ \to J/\psi p$ &  4.12 &
$P_{c1}^+ \to \eta_c p$ & 14.81 \\
\hline
$P_{c1}^0 \to J/\psi n$ &  4.11 &
$P_{c1}^0 \to \eta_c n$ & 14.78 \\
\hline
$P_{c1}^{s+} \to J/\psi \Sigma^+$ & 1.01 &
$P_{c1}^{s+} \to \eta_c \Sigma^+$ & 4.06 \\
\hline
$P_{c1}^{s0} \to J/\psi \Sigma^0$ & 1.00 &
$P_{c1}^{s0} \to \eta_c \Sigma^0$ & 4.04 \\
\hline
$P_{c1}^{s-} \to J/\psi \Sigma^-$ & 0.98 &
$P_{c1}^{s-} \to \eta_c \Sigma^-$ & 4.00 \\
\hline
$\tilde P_{c1}^{s0} \to J/\psi \Lambda^0$ &  7.13 &
$\tilde P_{c1}^{s0} \to \eta_c \Lambda^0$ & 26.76 \\
\hline
$P_{c1}^{ss0} \to J/\psi \Xi^0$ & 1.94 &
$P_{c1}^{ss0} \to \eta_c \Xi^0$ & 7.85 \\
\hline
$P_{c1}^{ss-} \to J/\psi \Xi^-$ & 1.90 &
$P_{c1}^{ss-} \to \eta_c \Xi^-$ & 7.75 \\
\hline
 \multicolumn{4}{c}{Second Family $P_{c2}$} \\
\cline{1-4}
\hline
$P_{c2}^+ \to J/\psi p$ & 39.82 &
$P_{c2}^+ \to \eta_c p$ &  5.45 \\
\hline
$P_{c2}^0 \to J/\psi n$ & 39.77 &
$P_{c2}^0 \to \eta_c n$ &  5.45 \\
\hline
$P_{c2}^{s+} \to J/\psi \Sigma^+$ & 10.99 &
$P_{c2}^{s+} \to \eta_c \Sigma^+$ &  1.58 \\
\hline
$P_{c2}^{s0} \to J/\psi \Sigma^0$ & 10.93 &
$P_{c2}^{s0} \to \eta_c \Sigma^0$ &  1.57 \\
\hline
$P_{c2}^{s-} \to J/\psi \Sigma^-$ & 10.84 &
$P_{c2}^{s-} \to \eta_c \Sigma^-$ &  1.56 \\
\hline
$\tilde P_{c2}^{s0} \to J/\psi \Lambda^0$ & 72.09 &
$\tilde P_{c2}^{s0} \to \eta_c \Lambda^0$ & 10.08 \\
\hline
$P_{c2}^{ss0} \to J/\psi \Xi^0$ & 21.25 &
$P_{c2}^{ss0} \to \eta_c \Xi^0$ &  3.06 \\
\hline
$P_{c2}^{ss-} \to J/\psi \Xi^-$ & 20.99 &
$P_{c2}^{ss-} \to \eta_c \Xi^-$ &  3.10\\
\hline
 \multicolumn{4}{c}{Third Family $P_{c3}$} \\
\cline{1-4}
\hline
$P_{c3}^+ \to J/\psi p$ & 80.04 &
$P_{c3}^+ \to \eta_c p$ &  $0.13 \, \beta^2$ \\
\hline
$P_{c3}^0 \to J/\psi n$ & 79.93 &
$P_{c3}^0 \to \eta_c n$ &  $0.13 \, \beta^2$ \\
\hline
$P_{c3}^{s+} \to J/\psi \Sigma^+$ & 22.45 &
$P_{c3}^{s+} \to \eta_c \Sigma^+$ &  $0.05 \, \beta^2$ \\
\hline
$P_{c3}^{s0} \to J/\psi \Sigma^0$ & 22.63 &
$P_{c3}^{s0} \to \eta_c \Sigma^0$ & $0.05 \, \beta^2$ \\
\hline
$P_{c3}^{s-} \to J/\psi \Sigma^-$ & 22.17 &
$P_{c3}^{s-} \to \eta_c \Sigma^-$ & $0.05 \, \beta^2$ \\
\hline
$\tilde P_{c3}^{s0} \to J/\psi \Lambda^0$ & 146.67 &
$\tilde P_{c3}^{s0} \to \eta_c \Lambda^0$ & $0.28 \, \beta^2$ \\
\hline
$P_{c3}^{ss0} \to J/\psi \Xi^0$ & 43.58 &
$P_{c3}^{ss0} \to \eta_c \Xi^0$ & $0.10 \, \beta^2$ \\
\hline
$P_{c3}^{ss-} \to J/\psi \Xi^-$ & 43.09 &
$P_{c3}^{ss-} \to \eta_c \Xi^-$ & $0.10 \, \beta^2$ \\
\hline
\end{tabular}
\label{results:charm}
\end{center}
\end{table}

We can use the upper limit on the branching ratio of the $P_c(4457)^+$ 
recently set by the GlueX Collaboration at JLab~\cite{Ali:2019lzf} to
constrain the size parameter $\Lambda$ of this LHCb pentaquark assuming 
spin-parity $J^P=\frac{3}{2}^-$ for this state. We do not consider $P_c(4312)^+$ 
and $P_c(4440)^+$ states in our estimate because they have spin-parity $J^P = \frac{1}{2}^-$
in our case, which is different from the assumption of the GlueX experiment. 
As pointed out by the GlueX Collaboration~\cite{Ali:2019lzf} 
their upper limits are sensitive to the spin-parity quantum numbers. 
In particular, the limits become by a factor of 5 smaller for the 
$J^P = \frac{5}{2}^+$ assignment.  
For an estimate of the branching of the $P_c(4457)^+$ state we use the central value  
for the decay width of the recent LHCb data analysis~\cite{Aaij:2019vzc}: 
\eq 
\Gamma_{P_c(4457)^+} = 6.4 \ {\rm MeV}\,. 
\en 
We find the following constraint: $\Lambda_{{P_c(4457)^+}} \le 145$ MeV. Note that the 
$\Lambda$ parameter characterizes the binding forces acting on a charmed baryon and a meson in 
the bound state of a hidden charm pentaquark. 

\begin{table}[htb]
\begin{center}
\caption{Two-body decay rates of hidden bottom pentaquarks in MeV  
[results for decay 
rates should be multiplied by factor $(\Lambda/1 \ {\rm GeV})^3$]} 

\vspace*{.1cm}

\def\arraystretch{1.}
\begin{tabular}{c|c||c|c}
\hline
   \ \ Mode \ \ 
&  \ \ Decay rate (MeV) \ \ 
&  \ \ Mode \ \ 
&  \ \ Decay rate (MeV) \ \ \\
\hline 
 \multicolumn{4}{c}{First Family $P_{b1}$} \\
\cline{1-4}
\hline
$P_{b1}^+ \to \Upsilon p$ &  0.38 &
$P_{b1}^+ \to \eta_b p$   &  1.20 \\
\hline
$P_{b1}^0 \to \Upsilon n$ &  0.38 &
$P_{b1}^0 \to \eta_n n$   &  1.20 \\
\hline
$P_{b1}^{s+} \to \Upsilon \Sigma^+$ & 0.12 &
$P_{b1}^{s+} \to \eta_n \Sigma^+$   & 0.38 \\
\hline
$P_{b1}^{s0} \to \Upsilon \Sigma^0$ & 0.12 &
$P_{b1}^{s0} \to \eta_n \Sigma^0$   & 0.38 \\
\hline
$P_{b1}^{s-} \to \Upsilon \Sigma^-$ & 0.12 &
$P_{b1}^{s-} \to \eta_b \Sigma^-$   & 0.38 \\
\hline
$\tilde P_{b1}^{s0} \to \Upsilon \Lambda^0$ &  0.76 &
$\tilde P_{b1}^{s0} \to \eta_n \Lambda^0$   &  2.38 \\
\hline
$P_{b1}^{ss0} \to \Upsilon \Xi^0$ & 0.24 &
$P_{b1}^{ss0} \to \eta_b \Xi^0$   & 0.76  \\
\hline
$P_{b1}^{ss-} \to \Upsilon \Xi^-$ & 0.24 &
$P_{b1}^{ss-} \to \eta_b \Xi^-$   & 0.76 \\
\hline
 \multicolumn{4}{c}{Second Family $P_{b2}$} \\
\cline{1-4}
\hline
$P_{b2}^+ \to \Upsilon p$ &  3.27 &
$P_{b2}^+ \to \eta_b p$   &  0.41 \\
\hline
$P_{b2}^0 \to \Upsilon n$ &  3.27 &
$P_{b2}^0 \to \eta_b n$   &  0.41 \\
\hline
$P_{b2}^{s+} \to \Upsilon \Sigma^+$ & 1.03 &
$P_{b2}^{s+} \to \eta_b \Sigma^+$   & 0.13 \\
\hline
$P_{b2}^{s0} \to \Upsilon \Sigma^0$ & 1.03 &
$P_{b2}^{s0} \to \eta_b \Sigma^0$   & 0.13 \\
\hline
$P_{b2}^{s-} \to \Upsilon \Sigma^-$ & 1.03 &
$P_{b2}^{s-} \to \eta_b \Sigma^-$   & 0.13 \\
\hline
$\tilde P_{b2}^{s0} \to \Upsilon \Lambda^0$ & 6.41 &
$\tilde P_{b2}^{s0} \to \eta_b \Lambda^0$   & 0.78 \\
\hline
$P_{b2}^{ss0} \to \Upsilon \Xi^0$  & 2.07 &
$P_{b2}^{ss0} \to \eta_b \Xi^0$    & 0.26 \\
\hline
$P_{b2}^{ss-} \to \Upsilon \Xi^-$  & 2.06 &
$P_{b2}^{ss-} \to \eta_b \Xi^-$    & 0.26 \\
\hline
 \multicolumn{4}{c}{Third Family $P_{b3}$} \\
\cline{1-4}
\hline
$P_{b3}^+ \to \Upsilon p$ & 6.57 &
$P_{b3}^+ \to \eta_b p$   & $0.004 \, \beta^2$ \\
\hline
$P_{b3}^0 \to \Upsilon n$ & 6.56 &
$P_{b3}^0 \to \eta_b n$   & $0.004 \, \beta^2$ \\
\hline
$P_{b3}^{s+} \to \Upsilon \Sigma^+$ & 2.07 &
$P_{b3}^{s+} \to \eta_b \Sigma^+$ &  $0.002 \, \beta^2$ \\
\hline
$P_{b3}^{s0} \to \Upsilon \Sigma^0$ & 2.07 &
$P_{b3}^{s0} \to \eta_b \Sigma^0$   & $0.002 \, \beta^2$ \\
\hline
$P_{b3}^{s-} \to \Upsilon \Sigma^-$ & 2.06 &
$P_{b3}^{s-} \to \eta_b \Sigma^-$   & $0.002 \, \beta^2$ \\
\hline
$\tilde P_{b3}^{s0} \to \Upsilon \Lambda^0$ & 12.87 &
$\tilde P_{b3}^{s0} \to \eta_b \Lambda^0$   & $0.009 \, \beta^2$ \\
\hline
$P_{b3}^{ss0} \to \Upsilon \Xi^0$ & 4.15 &
$P_{b3}^{ss0} \to \eta_b \Xi^0$   & $0.003 \, \beta^2$ \\
\hline
$P_{b3}^{ss-} \to \Upsilon \Xi^-$ & 4.13 &
$P_{b3}^{ss-} \to \eta_b \Xi^-$   & $0.003 \, \beta^2$ \\
\hline
\end{tabular}
\label{results:bottom}
\end{center}
\end{table}

In conclusion, we presented a calculation for the strong two-body decays 
of hidden charm and bottom pentaquarks into pairs of a light baryon and 
heavy quarkonia. We evaluate the final results in the heavy quark limit 
using only one free parameter -- the scale parameter $\Lambda$ characterizing 
the binding forces of the hadronic constituents in the hidden heavy pentaquark. 
Our predictions for the decay rates scale as $\Lambda^3$,
therefore the results are very sensitive to the choice of this parameter. 
Using recent data of the LHCb Collaboration~\cite{Aaij:2015tga}-\cite{Aaij:2019vzc} 
on the total widths of nonstrange hidden charm pentaquarks and of 
the GlueX Collaboration~\cite{Ali:2019lzf} on the upper limit of the partial 
two-body decay width of the $P_c(4457)^+$ state with $J^P = \frac{3}{2}^-$ we derive 
preliminary result for the upper limit for our scale parameter~$\Lambda$. Future and 
more precise experiments on the decay properties of hidden heavy pentaquarks can give 
strong constraints on the model parameter and insights into the predictions given 
in the last tables.

\begin{acknowledgments}

This work was funded by
the Carl Zeiss Foundation under Project ``Kepler Center f\"ur Astro- und
Teilchenphysik: Hochsensitive Nachweistechnik zur Erforschung des
unsichtbaren Universums (Gz: 0653-2.8/581/2)'',
by CONICYT (Chile) under Grant PIA/Basal FB0821 
and by FONDECYT (Chile) under Grant No. 1191103, 
by Tomsk State University Competitiveness
Improvement Program and the Russian Federation program ``Nauka''
(Contract No. 0.1764.GZB.2017). 

\end{acknowledgments}

\end{document}